\documentclass[12pt,preprint]{aastex}
\shorttitle{Survey of the Magellanic Clouds}
\shortauthors{Massey}
\begin{document}

\title{A {\it UBVR} CCD Survey of the Magellanic Clouds} 

\author{Philip Massey\altaffilmark{1}}
\affil{Lowell Observatory, 1400 West Mars Hill Road,
Flagstaff, AZ 86001}
\email{massey@lowell.edu}

\altaffiltext{1}{Visiting astronomer,
Cerro Tololo Inter-American Observatory, a division of the
National Optical Astronomy Observatory, which is operated by the 
Associations of Universities for Research in Astronomy, Inc.,
under cooperative agreement with the National Science Foundation.}

\begin{abstract}
We present photometry and a preliminary interpretation of a 
{\it UBVR} survey of the Large and Small Magellanic Clouds,
which covers
14.5 deg$^2$ and 7.2 deg$^2$, respectively.
This study is aimed at obtaining well-calibrated data on the brighter, 
massive stars, complementing
recent, deeper CCD surveys.  Our catalog contains 179,655 LMC and 84,995 SMC
stars brighter than $V\sim 18.0$, and is photometrically complete to
$U\sim B\sim V\sim 15.7$, and $R\sim 15.2$, although stars in crowded
regions are selectively missed.   We provide tentative
cross-reference between our catalog stars and the 
stars with existing spectroscopy.
Our photometry agrees well with the photoelectric work in {\it V} and 
{\it B--V},
and agrees well for {\it U--B} for the bluest stars, but we find a large
discrepancy (0.3~mag) in the {\it U--B} color at {\it U--B}$\sim 0.0$.  
Examination of the colors
of stars with known spectral types suggests that the problem {\it may} 
lie with
the photoelectric data.
We discuss the population
of stars seen towards the two Clouds, identifying the features in the 
color-magnitude diagram, and using existing spectroscopy to help
construct H-R diagrams.  We derive improved
values for the blue to red star ratios in the two Clouds, emphasizing the
uncertainties involved in this before additional spectroscopy. We compare
the relative number of RSGs and Wolf-Rayet stars in the LMC and SMC with that
of
other galaxies in the Local Group, demonstrating a very strong, tight trend
with metallicity, with the ratio changing by a factor of 160 from the SMC
to M31. We also
reinvestigate the initial mass function of the massive stars found outside
of the OB associations.  With the newer data, we find that the IMF slope
of this field population is very steep, with $\Gamma\sim -4\pm0.5$, 
in agreement with our earlier work. This is in sharp contrast to the IMF
slope found for the massive stars with OB associations ($\Gamma\sim -1.3$).
Although much more spectroscopy is needed to make this result firm,
incompleteness can no longer
be invoked as an explanation.

\end{abstract}


\section{Introduction}

The Magellanic Clouds (MCs) are the Milky Way's nearest neighbors, and as
such serve as useful laboratories for studies of stellar astrophysics.
Objective prism surveys in the 1960s and 1970s (Sanduleak 1969a, 1969b,
1975; Brunet et al.\ 1975;  Azzopardi \& Vigneau 1975, 1979; 
Rousseau et al.\ 1978) helped
identify many of the less crowded early-type members of the Clouds.
These surveys produced two catalogs which still serve as the primary
references: a list of 1822 LMC members by Rousseau et al.\ (1978), and
a list of 524 SMC members by Azzopardi \& Vigneau (1982).  These catalogs
represent a small fraction of the overall number of massive stars in
the Clouds, 
as they are generally incomplete in the more crowded regions
where most of the OB stars lie, but they do serve as important sources
of data for the field stellar content of the Clouds (Massey et al.\ 1995;
Massey 1998a).  Subsequent slit
spectroscopy of hundreds of these stars yielded H-R diagrams with
sufficient accuracy to provide useful tests of current massive star
evolution models (Fitzpatrick \& Garmany 1990; Massey et al.\ 1995).

However, good photometry has lagged behind the spectroscopic efforts.
Although many small regions have been imaged and studied with CCDs,
the general field population of the Magellanic Clouds has only
photoelectric photometry, typically
conducted with large (18 arcsec diameter)
apertures (Isserstedt 1975, 1978, 1979, 1982; Azzopardi \& Vigneau 1975).
Such photoelectric photometry is limited by both stellar and nebular
contamination.  Our motivation in improving this situation comes about
in part by our Cycle 9 {\it HST} program to find physical parameters
for hot stars in the Magellanic Clouds.  For modeling, good knowledge
of $M_V$ is needed in order to estimate the stellar radius, and this
requires good understanding of both $V$ and the color excess $E(B-V)$.
Accurate, well-calibrated $U-B$ photometry is useful for
selecting the earliest type stars for spectroscopic studies with the end
goal of understanding the stellar initial mass function (IMF) of the Clouds, an
issue resolved for the stellar associations but not for the field content
(Massey 1998a, Parker et al.\ 1998). In addition, the identification
of the red supergiant (RSG) population of the Magellanic Clouds would
serve an important constraint on stellar evolutionary theory, and for
this, $R$ is crucial:  Massey
(1998b) has demonstrated the usefulness of {\it BVR} two-color diagrams
in separating bona-fide RSGs from foreground dwarfs.  Finally, we note
that a {\it UBVRI} photometric survey 
of more distant Local Group galaxies currently forming stars is underway
with the KPNO and CTIO 4-m telescopes and Mosaic cameras; 
galaxy-wide CMDs of the Magellanic Clouds will
serve as a useful comparison to these data.

Accordingly we undertook a survey of the LMC and SMC using
the CCD system of the Michigan Curtis Schmidt
telescope at CTIO.  The goal was to provide good ($<3$\%) photometry at
{\it UBVR} over the range of interest to massive star work, e.g., roughly
$V\sim 12.5$ to 15, with care being taken not to saturate the brighter
stars of interest, and to provide excellent calibration particularly at
$U-B$. The resulting catalog contains 179,655 LMC and 84,955 SMC stars
brighter than $V\sim 18.0$.  We realized that this work would be useful of
others engaged in massive star studies in the Clouds, and would serve as
a useful complement to other contemporary surveys of the Magellanic Clouds: the
\anchor{http://www.ctio.noao.edu/mcels/} {Magellanic Cloud Emission-line
Survey} of Smith et al. (1999), the {\it IJK} Deep Near Infrared Survey
(Cioni et al.\ 2000), 2Mass {\it JHK} survey (Van Dyk \& Cutri 1999),
and the {\it UBVI} Magellanic Clouds Photometry Survey (Zaritsky, Harris,
\& Thompson 1997; Zaritsky et al.\ 1999).  The latter covers
a larger area, goes considerably deeper (50\% complete at $V\sim 21$),
and is obtained at higher resolution; however, our survey provides better
data for the brighter stars, higher precision calibration (particularly
at {\it U}), and complements that survey by including
{\it R}.

In what follows, we adopt distance moduli of 18.48 and 18.94 for the LMC and
SMC, respectively (Westerlund 1997, van den Bergh 2000), and average
reddening values $E(B-V)=0.13$ and 0.09 (Massey et al.\ 1995).

\section{Construction of the Catalog}

\subsection{Observations and Reductions}
All of the data were taken with the Curtis Schmidt telescope at CTIO using
a Tektronix 2048$\times$2048 CCD with 24$\mu$m pixels.  The scale was
2.32 arcseconds/pixel, and each image covered $1.3^\circ \times 1.3^\circ$.
We used the ``Harris set" filters which provides relatively 
straight-forward transformations to the standard {\it UBVR} system
(Massey et al.~2000a)

Our data set consists of 6 SMC and 11 LMC fields.  We illustrate our
coverage in Figs.~1 and 2.  The total area uniquely covered is
14.5 deg$^2$ and 7.2 deg$^2$ in the LMC and SMC, respectively.

Observations of the 6 SMC fields and 8 of the 11 
LMC fields were taken on the photometric
night of (UT) 8 January 1999, with standards drawn from Landolt (1992).
However, only 10 standards from two pointings were made, and we 
were not satisfied with the color range, as there were no stars with
$U-B<0.0$.  Accordingly, we reobserved all 8 of the LMC fields, plus
an additional 3, during four photometric nights (UT) 2001 March 28-30 and
April 1.
During the 2001 
run, 194 standards were observed in 51 separate pointings, with care being
taken each night to bracket the range of airmasses of the LMC field.
Standard fields included some of the larger ``selected areas" 
of Landolt (1992), allowing standards to be measured in a variety of
location in the field, as a check on the flat-fielding and spatial
uniformity of the color terms.
The standards covered the following range in color: $-1.1<U-B<2.0$,
$-0.3<B-V<2.3$, $-0.14<V-R<1.4$.

The images are undersampled, with full-widths at half-maximum of
0.9-1.3 pixels.  We measured all of the standards using a digital aperture 
of radius 3.5 pixels (8.1 arcsec).

Our transformation equations were
of the form: 

$$u=U+u_1+u_2\times X+u_3\times (U-B)$$
$$b=B+b_1+b_2\times X+b_3\times (B-V)$$
$$v=V=v_1+v_2\times X+v_3\times (B-V)$$
$$r=R=r_1+r_2\times X+r_3\times (V-R)$$

\noindent
where U, B, V, and R denote the standard magnitudes, {\it u, b, v,} and {\it r} denote
the instrumental magnitudes, $X$ is the airmass, and $u_1$ through $r_3$ 
are 
coefficients to be determined.  
We reduced each night separately, but found
(as expected) that the color terms were constant, and re-reduced
the 2001 data with the color terms fixed for all four nights.
The values are given in Table~1.  The values for the extinction, and
colors terms, are quite typical.  The RMS of the fits were typically
0.025~mag for {\it U}, and 0.015~mag for {\it B, V,} and {\it R}.

Despite the considerable range in color, we obtained satisfactory fits
using a linear color term {\it except}
at $U-B$.  As is our experience with {\it all} CCDs used with
this type of U filter, two separate slopes are needed for the color term,
depending upon whether $U-B$ is greater than or less than 0.0.  

The SMC and two LMC data sets were then measured in a similar manner:
IRAF/daophot was used to perform aperture photometry with a digital
aperture of 3.5-pixel radius of all unsaturated stars that were more than
8-10$\sigma$ above the noise.  After applying the relevant transformations
to the two runs, we found good agreement in $V$, $B-V$, and $V-R$, but the
$U-B$ data deviated for the bluest stars, and we applied a correction to
bring the 1999 data into accord with the 2001 data, which were the
better calibrated set.

Celestial coordinates were computed for each field using the Space Telescope
Guide Star Coordinate (GSC) system; the LMC fields are tied to the GSC 
solution for plate-id 06B0 and 0186,
while the SMC fields are tied to the GSC solutions for plate-id 06B1.
In constructing the catalogs, photometry from adjacent
fields were averaged; the agreement between these fields were checked,
and found to be good.  We insisted that a star be identified in
{\it B and
V} and {\it either} $R$ {\it or} $U$ to be included in our catalog.

\subsection{Errors and Completeness}

We have tabulated our average photometric errors in Table~2.  In order
to understand what these errors mean in terms of the science we plan to
achieve, we remind the reader that our goal was to obtain good photometry
for the {\it most massive} stars in the Magellanic Clouds, and in
particular stars with masses $>25\cal M_\odot$.  In the SMC
a {\it zero-age main-sequence} (ZAMS) star with mass 25$\cal M_\odot$
will be found with $V\sim15.4$, and will brighten to $V\sim14.3$ after
5~Myr of evolution. 
On the ZAMS such a star will be of spectral type
O8~V, and after 5~Myr will be of O9.5~III; see Massey (1998a).  Thus,
allowing for reddening, over this same period a 25$\cal M_\odot$ star
would evolve from $U\sim 14.1$ to $U\sim 13.1$.  Of course, there
are regions of the Clouds where the reddening is significantly higher;
even in our sample of interesting early-type O stars we have examples
where $E(B-V)\sim0.6-0.8$ in the LMC.  Our $25\cal M_\odot$ star would
be found at $V\sim 17$ in the LMC if the reddening were extreme.

Completeness is harder to evaluate.  In the crowded regions of the
Clouds our catalog will be incomplete even at bright magnitudes,
given our spatial sampling. 
In terms of completeness with magnitude, we
show in Figs.~\ref{fig:LMChists} and \ref{fig:SMChists} histograms of
the number of stars as a function of magnitude for all four filters.
Certainly the data becomes increasing incomplete as it approaches the peak
of the distributions.  The {\it peaks} in the distributions of the number
of stars occur at roughly $V\sim16.2$ and $B\sim16.2$, and we therefore
expect significant incompleteness beyond $V\sim 16$ and $B\sim 16$.
In addition, we have included only stars which were measured at {\it U
or R} in addition to {\it B and V}.  The $U$ distribution peaks
at $\sim16.2$ as well.  The exposures at $R$ were shorter so as to not
saturate bright red supergiants, and here the distribution peaks about
half a magnitude brighter, at $R\sim 15.7$.  We expect that our completeness
limit is several bins below the peaks, and we therefore estimate that our
data are photometrically complete to $U\sim B \sim V \sim 15.7$, and $R\sim 15.2$.  Thus, we only just meet our goal of completeness to a ZAMS star with
25$\cal M_\odot$. We note that 
crowding is an additional concern, and that stars in crowded
regions are very poorly represented in our catalog.  We also note that stars
brighter than $V\sim12$ will be saturated; fortunately stars brighter than
this in the Magellanic Clouds are few and well studied.

\section{The Catalog}
We present our LMC and SMC catalogs in Table~3.  We include the
coordinates of a star, its photometry (99.99 denoting no measurement),
and the errors, along with the number of times a star was measured in
each filter.  We have sorted the stars by right ascension, and assigned
a running number as our catalog designation. 
We have truncated each catalog at $V=18$, near the
plate-limit, nearly two magnitudes below the turn-over in the number
of stars.  The final LMC catalog contains 179,655 stars, of which
62\% have $U$ measures, 74\% have $R$ measurements, and 36\% have both.
About 19\% of these stars were observed on more than one frame.
The SMC catalog contains 84,955 stars, and has similar statistics:
70\% have $U$ measurements, 75\% have $R$ measurements, and 45\% have
both $U$ and $R$.  There is more overlap among the SMC frames, and the
fraction of stars observed more than once is 45\%.

\subsection{Cross-IDs: MC Stars with Known Spectral Types}

Many previous catalogs of bright LMC and SMC members have been published.
For the LMC, the usual starting point is the objective prism survey of 
Sanduleak (1969a), supplemented by Brunet et al.\ (1975).  Early slit
spectroscopy most notably 
by Feast, Thackeray, \& Wesselink (1960), Ardeberg et al.\ (1972),
Brunet et al.\ (1973) were then included in the 
Rousseau et al.\ (1978) catalog, which contained MK types for 266 stars.
The Rousseau et al.\ (1978) catalog also gave 
new (objective prism) spectral types and $V$ photometry. 
Fitzpatrick \& Garmany (1990) updated this catalog to included
the Isserstedt (1979, 1982) and well-determined spectral classification
for $\sim 300$ OB stars.  This catalog was subsequently updated by Massey
et al.\ (1995), who gave new spectral types for another $\sim 200$ LMC stars,
including those in two ``incompleteness fields".
Since that time additional spectral types have been 
provided by Massey, Waterhouse, \& DeGioia-Eastwood (2000b), among others.
In Table 4 we present the cross-identification between our stars and those
from the updated Rousseau et al.\ (1978) catalog.  

Many recent studies of OB associations in the LMC have presented good
spectral types and CCD photometry, and we do not repeat those data here.
Although Table~3 contains many stars associated with OB associations, they
are more complete in the {\it field}, where uncertainties still exist over
membership and origins (see discussion in and following Massey 1999).
We call attention to the LMC ``incompleteness fields" stars given by 
Massey et al.\ (1995).  Unfortunately, due to an error by the first author,
the list of stars and spectral types contained incorrect coordinates, making
their identification impossible.  We rectify this mistake in Table~5,
where we repeat the photometry and spectral types but with corrected
positions. We are indebted to Nolan Walborn, who first called this error to
our attention back in 1996.

For the SMC, the objective prism work of Azzopardi \& Vigneau (1982)
remains the definitive catalog of probable SMC members.   
The Azzopardi \& Vigneau (1982) catalog
contained spectral classifications by either objective prism or slit
spectroscopy; to this, Garmany, Conti, \& Massey (1987) provided 
slit spectrograph MK classifications for 120 OB stars.  In addition,
Massey et al.\ (1995) provided $\sim 90$ types for other SMC stars,
including those in the incompleteness field.  We provide cross-reference
to the stars with existing spectroscopy in Table 6, where we also include
the ``missing" SMC stars discussed by Azzopardi \& Vigneau (1982): twenty
stars from the list of Feast et al.\ (1960), Thackeray (1962), and
Sanduleak (1968), and Sanduleak (1969b), which were all outside the
region resurveyed by Azzopardi \& Vigneau.
We have included revised spectral types for
some SMC stars by Massey et al.\ (2000b).
The ``western extremity" region looked at by Sanduleak (1975) is outside
our survey area.
 
We note that many of the OB associations of the SMC have also been
studied with both CCD photometry and spectroscopy, and we do not repeat
those data here.  

In preparing the cross identifications the sheer volume of the data forced
us to rely upon the published positions to look for reasonable matches.
Since the published positions are given only to 1 arcmin in the primary
literature, this process could not be completely accurate. For a few
dozen ambiguous identifications, we checked our fields with the published
finding charts by eye, and found no problems in the areas immediately
surrounding the object we were checking.  Nevertheless, we are certain there
are omissions in Tables 4 and 6, and there may be erroneous identifications
as well.  We urge users interested in the photometry of their favorite star
to extract accurate coordinates via the laborious process of comparing the
published finding charts with the Digitized Sky Survey, and then using the 
latter to produce accurate coordinates.  Those coordinates should match those
in Tables 4 and 6 to within an arcsecond.  If not, the wrong identification
may have been made.

\subsection{Comparison of our Photometry with that of Others: Is There a 
Problem?}

With the cross-identifications above, we can now compare our CCD photometry
with the published photoelectric values.  Our digital measuring aperture
is in fact not much smaller than that used for the vast majority of
published photoelectric photometry (16.2 vs.\ 18 arcsec), and so one might
expect our photometry to be comparably affected by neighboring stars.
However, CCD photometry has two intrinsic advantages over the photoelectric
work: (1) our stellar photometry
is not affected by centering errors within the aperture,
and (2) our sky measurements are local and free of the effects of resolved
stars. Our own experience in attempting to do photoelectric photometry
in the Clouds suggests that both of these effects can be large.

We compare our photometry to the photoelectric work in Fig~\ref{fig:photdifs}.
There is little difference in $V$ or $B-V$ for the LMC, with the 
average differences of $\Delta V=0.000\pm0.002$~mag and 
$\Delta B-V=+0.001\pm0.002$~mag 
(in the sense of CCD {\it minus} photoelectric).
The differences for $V$ and $B-V$ are larger for our SMC data:
$\Delta V=-0.07\pm 0.004$ and $\Delta B-V=+0.031\pm 0.002$. 
Although the SMC catalog relies upon the less well calibrated 1999
data, we note that there is no systematic difference between the
2001 and 1999 {\it LMC} data, and the LMC data {\it does} agree well with
that from the literature.  Could there be a systematic
problem in the published photoelectric photometry of stars in the SMC
by nearly a tenth of a magnitude?  We consider this issue unresolved.

Most interesting is the systematic difference in the $U-B$ data evident
in Fig~\ref{fig:photdifs}.  In both the LMC and the SMC, we see excellent
agreement between our photometry and the photoelectric   
$U-B$ data for the bluest stars ($U-B\sim -1$).  However, there a 
systematic trend with $U-B$ color, with the difference between the
two systems amounting to 0.3~mag by $U-B\sim 0$.  We were at first convinced
that something bizarre had transpired in our transformation equations.
However, we had plenty of well-calibrated standard stars over this
interval (and indeed from $U-B=-1.1$ to $+2.0$), and after exhaustive
investigation to find the source of the problem we began to entertain
the possibility that there was something wrong with
the published photoelectric
work in this color range. 

We consider an external check.  Table~7
lists the stars with slit spectral types for stars with $U-B>-0.3$, i.e.,
A and early F-type supergiants.
We provisionally adopt the {\it intrinsic} colors 
corresponding to these
spectral types based on the FitzGerald (1970) calibration of Galactic
stars.  We can then determine the color excess 
$E(B-V)=(B-V)_{\rm ccd}-(B-V)_o$.
Based on this, we can then calculate the $(U-B)_o$ expected on the basis
of our photometry with the assumption that $E(U-B)/E(B-V)=0.72$, and compare
that with that expected from the FitzGerald (1970) calibration.  We see
that the $\Delta (U-B)_o = (U-B)_{o\rm (Photometry)}-(U-B)_{o\rm (Spectral Type)}$
averages to $0.03\pm0.01$ for our CCD data, but that $\Delta (U-B)_o$
averages to $0.28\pm0.02$ for the photoelectric data. 
Thus it would appear that the problem {\it may} be with the photoelectric work.

The photoelectric photometry referenced in Azzopardi \& Vigneau (1982) and
Rousseau et al.\ (1978) come from a variety of seemingly different sources,
with good agreement.  For instance, the SMC photoelectric
photometry we list in Table~7 is derived from Isserstedt (1978) and 
Azzopardi \& Vigneau (1979).  Similarly the LMC photoelectric photometry
in Table~7 is derived from three separate sources.  However, closer inspection
reveals that all five of these sources used the same equipment, at the same
observatory, employing the same standards.  We do see a hint of this problem
in the two-color diagram shown Figure 11B of Azzopardi \& Vigneau (1979).
There are plenty of A-type stars which lie $\sim 0.3$~mag above the intrinsic
relationship for supergiants, and the smattering of stars that fall on
the supergiant line could actually be dwarfs.  We do note that this problem
is not characteristic of {\it all} photoelectric photometry of the Clouds.
For instance, the A-type supergiants classified by Humphreys (1983) have
photometry that is carefully described by
Ardeberg \& Maurice (1977) or Ardenberg et al.\ (1972), and
their dereddened colors agree very well with that expected on the basis
of their spectral types.

We would prefer to believe that it is
our own CCD photometry that has a calibration error, and perhaps this is the
case.  However, the data in Table~7 do suggest that the dereddened CCD
colors agree better with the expected intrinsic colors than do the
photoelectric data.
The only other possible explanation is that
there is a coincidental difference of 0.3~mag in the intrinsic $(U-B)_o$ color
of A-type supergiants in the Magellanic Clouds, compared to those in the
Milky Way, due to metallicity effects.  We argue that this is unlikely on
two grounds. First, there is little difference in the $\Delta (U-B)_o$ given
in Table~7 between the SMC and the LMC, where the difference in metallicity
is as significant as between the LMC and the Milky Way.  Secondly, we can
use the model atmosphere calculations to explore the magnitude of the effect
we would expect.  Using the Kurucz (1992) ATLAS9 models, we find that 
$U-B$ should change by $<0.01$~mag over the metallicity range from the
Milky Way to the SMC.

\section{The Population of Stars Seen Towards the Magellanic Clouds}

In this section, we discuss the population of stars revealed in our catalog,
with the emphasis on distinguishing {\it bona fide} MC members from foreground
Galactic stars.

Figs.~\ref{fig:lmccmd} and \ref{fig:smccmd} show the 
$B-V$, $V$ color-magnitude
diagrams for the two Clouds.  We show the data with every point plotted, and
for additional clarity, with only 10\% of the data plotted. 
We see that there are generally three peaks of the brighter
stars: a blue clump peaking at $B-V=-0.1$, a broad peak extending
from $B-V\sim 0.4$ to 1.3, and a red clump moving diagonally to peak at
$B-V\sim 2.0$ at $V\sim 12$. 
These features are readily identified.  The reddenings in the LMC and the
SMC are small and typically uniform (Massey et al.\ 1995).
We show
the (de)reddening vectors corresponding to these values in the two
figures; for the purposes of discussing the coarse features of the CMDs
we will ignore reddening.

The ``blue plume" consists primarily of H-burning massive stars, the
dominant stellar population at the bright end of the CMD.  As a massive
star evolves, it will become brighter visually as the peak of the spectral
energy distribution shifts due to cooler effective temperatures.  Even so,
the stars with initial masses $>20\cal M_\odot$  end their H-burning lives
as early B-type supergiants (see Table 1 of Massey 1998a).  The very
brightest members of the ``blue plume" are in fact $15\cal M_\odot$
A-type supergiants; see Fig. 1 in Massey et al.\ (1995).

The very red,  strong, diagonal feature running from $B-V=1.5$ and
$V=14$ to $B-V\sim 2$ and $V=12$ consists of red supergiants (RSGs),
which are one type of He-burning massive stars.  At $B-V=1.5$ these
will be early K-type supergiants, but at $B-V=2$ these will be mid or
late M-type supergiants.  Since the bolometric correction changes by
2--3~mag over this range (Humphreys \& McElroy 1984), the brighter red
stars may be as much as a factor of 100 more bolometrically luminous than
the fainter RSGs.  These CMD features may be compared to the photographic
study of M33 by Humphreys \& Sandage (1980; see their Figs.~17-19) and by
Ivanov, Freedman, \& Madore (1993; see their Fig.~1).  The RSG sequence
appears to be much more strongly populated in the Magellanic Clouds.
Metallicity will affect the relative number of red supergiants (see
discussion in Massey 1998b), and the metallicity in M33 is higher than
in the Clouds, particularly in the inner parts.  Values for log O/H+12
vary from 8.75 in the inner part of M33 to 8.30 in the outer part
(Garnett et al.\ 1997),  while good values for the SMC and LMC are 8.13
and 8.37, respectively (Russell \& Dopita 1990).  The RSG branch of
the Magellanic Clouds are even more pronounced when compared to that of
relatively metal-rich galaxy M31 (Fig.~5 of Berkhuijsen et al.\ 1988),
where log O/H+12=9.0 (Zaritsky, Kennicutt, \& Huchra  1994).

The middle clumping of bright stars, ranging from B-V=0.4 to 1.2,
are dominated by foreground Galactic disk stars.  We can investigate
this quantitatively by using an updated version of the
Bahcall-Soneira model (Bahcall \& Soneira 1980).  Heather Morrison
was kind enough to provide the computer code that had been revised by
herself and Gary DaCosta.  Using reasonable assumptions, we computed the
number of Galactic stars expected to be seen per magnitude/color bin at
the galactic latitude and longitudes of the Small and Large Clouds.
We include in Figs~\ref{fig:lmccmd} and \ref{fig:smccmd} the results
of comparing these counts to what we find in our catalog.  Below $V=16$
we consider our data to be incomplete, but outside the region indicated
the models predict the foreground contamination to be negligible.

Interestingly, we can resolve the central foreground clump into two
components, a heavily populated one at $B-V\sim0.5$, and a more sparsely
populated one at $B-V\sim 1.0$.  The model calculations provide a clue:
the bluer, more heavily populated clump is dominated by disk dwarfs, while
the redder, more sparsely populated clump is dominated by disk giants.
The Milky Way's halo does not contribute substantially to the CMDs at
these magnitude levels.

\subsection{Blue Stars}
\label{sec:blue}

Essentially {\it all} of the blue stars are members of the
Magellanic Cloud; a galactic A0~V star would have to be at a distance of
2.5--16~kpc to be in our magnitude range of 12 to 16. 

For the bluest stars, the reddening-free parameter $Q$ provides a good
clue to the nature of the star.  Adopting a
standard reddening ratio $E(U-B)/E(B-V)=0.72$, we construct
$Q=(U-B)-0.72\times (B-V)$. For stars with with $Q<-0.4$ this provides
a unique mapping to $(B-V)_o$; above this value there is overlap between early-type stars
and yellow dwarfs.

We have used our photometry to compute the 
physical parameters $T_{\rm eff}$
and $M_{\rm bol}$ for the bluest stars in Tables 8A and 8B.  For the stars
with slit spectroscopy we have converted to $T_{\rm eff}$ using the
calibration of Vacca, Garmany, \& Shull (1996) or Humphreys \& McElroy (1984).
The correction for reddening is then made using the intrinsic colors corresponding to the spectral types, and the bolometric correction computed
on the basis of $T_{\rm eff}$.  For stars without spectral types, we have
estimated the physical parameters using the transformation equations given
by Massey et al.\ (2000b).  However, as
Massey (1998c) has emphasized, transformations via the photometry
are very useful for selecting stars interesting
for spectroscopy, but by themselves do not allow accurate parameters to be
determined.  We have restricted the entries in Tables 8 to stars
with a computed $M_{\rm bol}<-7.0$, corresponding to $20 \cal M_\odot$ on the ZAMS.

\subsection{Yellow Supergiants}
\label{sec:yellow}

We have found that most of the middle of the CMD is contaminated by
Galactic foreground stars.  However, over a limited range in
color, we should be able to separate yellow supergiants from foreground
dwarfs.  We illustrate this in Fig.~\ref{fig:2color}, where we show
the intrinsic $(U-B)_o$ and $(B-V)_o$ color for dwarfs and supergiants.
Given the typical (low) reddening,
we thus expect good
separation from $(B-V)_o = 0.25$ to 0.65, corresponding to mid-to-late F-type
supergiants. 

We select candidate F-type supergiants as follows: first, we determine
the approximate $(B-V)_o$ color using the observed colors and the
average $E(B-V)$ value for each cloud, selecting stars with
$(B-V)_o=0.26$ (roughly F5~I) to 0.68 (roughly F9~I).  A typical
G0~Ib star would have $M_V=-4.5$ (Humphreys \& McElroy 1984), 
and so we also require 
$M_V<-4.5$.  For the SMC, this corresponds to $V<14.7$, and for the
LMC (closer, but with higher reddening), $V<14.4$.  

\subsection{Red Supergiants}
\label{sec:red}

The identification of red supergiants in the Magellanic Clouds is of 
interest in understanding the evolution of massive stars.  
Massive stars evolve to red
supergiants (RSGs) and Wolf-Rayet stars (W-Rs) depending upon their
luminosities (mass) and composition. Evolutionary models predict that the
relative proportions of RSGs and W-Rs, and of RSGs and blue supergiants,
will increase at lower metallicities, a result of lower mass-loss;
see, for example Maeder, Lequeux, \& Azzopardi (1980).  
Recent studies have suggested that at high metallicities (e.g.,
log O/H+12=9.0, characteristic of M~31) there is a significant
paucity of higher luminosity RSGs compared to lower luminosity RSGs,
but there are {\it some}, suggesting that high mass stars become
RSGs albeit briefly even in M~31.

However, 
any sample of red stars in the right magnitude and
color range to be RSGs will be contaminated by foreground Galactic dwarfs
and more distant disk and halo giants.  We can estimate the degree
of contamination using the Bahcall \& Soneira (1980) model, as we did
in constructing Figs.~5 and 6 earlier.  For stars with $B-V>1.2$ and
$V<15$ we would expect about a 10\% contamination.  

Massey (1998b) explored these issues in regards to NGC~6822, M~33, and
M~31, and found that a $B-V$ vs.~$V-R$ two color diagram allowed one to
separate dwarfs from supergiants.  At a given $V-R$ color a low gravity
star will have larger $B-V$ value than a higher gravity star, due to the
increased importance of line blanketing at lower surface gravities,
which is most pronounced in the $B$ band due to the multitude of
weak metal lines in the region.  Good separation is expected even for
the lower metallicity of the SMC.

In Fig.~\ref{fig:rsgs} we show the $V-R$ vs.\ $B-V$ color-color plot.
The solid line denotes the expected separation of RSGs and foreground
stars, based upon $$(B-V)_{\rm cut}=1.599\times (V-R)^2+4.18\times (V-R)-1.04$$
For the SMC, the fractional contamination is 
14\%; in the case of the LMC, it is 17\%, comparable to what we expect
from the models.  We have used as our cut-off
$M_V=-4.0$, corresponding to $V=15.0$ (LMC) and 15.2 (SMC), magnitudes
for which we should still be complete.

We list in Table~10 the reddest stars seen towards the Magellanic Clouds.
We have used $V-R$ to compute the effective temperatures and bolometric
luminosities using the transformation equations of 
Oestreicher \& Schmidt-Kaler
(1999):  $$T_{\rm eff}=2710.+5342\times e^{-1.819(V-R)_o}$$ and
$${\rm BC}=1.688 - 3.103\times (V-R)_o-0.151\times (V-R)_o^2.$$
In order to determine $(V-R)_o$ we adopt the average reddening values.
Since $E(V-R)=0.53\times E(B-V)$, we expect $E(V-R)\sim 0.05$ on the
average.
We have included stars with 
$V-R>0.65$ in constructing this table 
[i.e., $(V-R)_o>0.60$] so as to include both K-type and M-type supergiants.
We have restricted the sample to $M_{bol}<-7.0$.
Thus Table~10 should be complete for all the luminous
K and M supergiants within our survey field.

A number of red stars have been classified in the Clouds; in particular,
see Humphreys (1979) and Elias, Frogel, \& Humphreys (1985).  
We have cross-referenced those stars in Table~10, using the identification
and coordinates given for the LMC by
Sanduleak \& Philip (1977) and for the SMC by Sanduleak (1989).  We have
not included cross-reference to the objective prism classifications
in these two papers, nor to the objective prism study of Prevot et
al.\ (1983), as we consider these primarily {\it candidate} RSGs.

We find that the two-color scheme for separating foreground stars works
well with the available SMC data, but poorly with the LMC data, as the
photometry would suggest that a number of the stars classified by
Humphreys (1979) as RSGs are actually foreground objects.  Additional
spectroscopy of suspected foreground and RSGs is needed to investigate
this further.

In determining the physical parameters of these stars (Table~10) we
emphasize that we used only the photometry, not the spectral types.
For red supergiants, the spectral classes depend upon the absolute
strength of spectral lines, and therefore we expect that metallicity
will play an important effect in addition to effective temperature.
This differs significantly from the situation with early-type stars,
where the classification scheme is based upon the relative strengths
of different ionization states of the same metal, i.e., He~I vs.\ He~II for
the O-type stars and Si~III vs.\ Si~IV for the early B-type stars, and
hence is metallicity independent to zeroth order.

\section{Preliminary Results and Discussion}

\subsection{H-R Diagrams}

We can use the data in the previous section to construct H-R diagrams for
the available data; we emphasize that only a limited subset of the data
as yet have slit spectral types, and that the resulting placement on the
H-R diagram is therefore uncertain.  We show the HRDs in Figs.~\ref{fig:lmchrd}
and \ref{fig:smchrd}, where we have separated the stars with spectral 
information from those with only photometry.  The figures may be compared to
Figs.~4 and 5 in Massey et al.\ (1995); note that in the present diagram
we exclude stars with $(B-V)_o>0.14$ except for the reddest stars,
with $(V-R)_o>0.60$.

\subsection{Blue to Red Ratios}

The red-to-blue supergiant ratio has long been recognized as varying
from galaxy to galaxy, and within a galaxy.
Van den Bergh (1973) suggested that this variation is due to metallicity
effects on stellar evolution.
Brunish, Gallagher, \& Truran (1986) found
satisfactory agreement between their stellar evolutionary models and
the data on blue and red supergiants 
compiled by Humphreys \& McElroy (1984), once corrections for selection
effects were made.  Langer \& Maeder (1995) have commented that ``no 
assumption on [sic] stellar model physics explored so far" properly 
reproduce
the ``observed" B/R ratio.  We have put observed in quotes because the data
quoted in such studies is often highly questionable.  Such studies typically
use the relative number of blue stars to red stars restricting
the sample to 
$M_{bol} < -7.5$, in order to achieve completeness (Humphreys \& McElroy 1984)
and to
remove the contamination of intermediate-mass
AGBs, which will affect the sample at $M_{\rm bol}>-7.1$ (Brunish et al. 1986).
There are two significant problems with these ``observed" ratios, however.
\begin{enumerate}
\item Prior photometric catalogs have {\it not} been complete to 
$M_{\rm bol}\sim-7.5$.  On the ZAMS, this corresponds to a mass of 
25$\cal M_\odot$, and a $\log T_{\rm eff}\sim 4.60$, for which the
bolometric correction is roughly $-3.8$mags.  Thus with typical reddenings
($E(B-V)=0.13$ for the LMC and 0.09 for the SMC), we would expect such a
star to be found at $V=15.2$ in the LMC, and $V=15.4$ in the SMC.
Azzopardi \& Vigneau (1982) consider their SMC catalog 80\% complete at
$B\sim 14.3$.
Such stars are only marginally within the completeness limit of the current
catalog!

\item The conversion to bolometric magnitudes of
extremely blue, or extremely red stars, from photometry alone is highly
uncertain, due to the very steep bolometric corrections at both the
blue and the red end.  In practice, this can be done reliably only with
good spectral types, and even then there is reliance upon stellar 
atmosphere models to predict how the bolometric corrections vary with
effective temperatures for differing metallicities.  The data in this
paper serve, we hope, as an encouragment for such spectroscopy.

\end{enumerate}

We use the data in the preceeding sections here to estimate the 
``blue to red" supergiant ratio in the SMC and the LMC.  For the
blue stars, we restrict ourselves to stars of spectral types O-A,
with the additional restrictions that $M_{\rm bol}<-7.5$, and $(B-V)_o<0.14$.
For the red stars, we again include only those with $M_{\rm bol}<-7.5$.
It is of interest, we feel, to consider the blue to red ratio based upon
including only the M-type stars, but also including stars of both K and M
type.  Thus, we have used the color criteria of $(V-R)_o>0.60$ and
$(V-R)_o>1.0$.  The resulting ratios are quite different, as is shown
in Table~11.  We remind the reader that the late-type supergiants are
of earlier type in the SMC than in the LMC; in particular, Elias et al.\ (1985)
find that the {\it median} type is earlier than M0 in the SMC, and M1 in the
LMC.  So, including potential K supergiants would appear to be necessary
if the blue to red ratios are to have any meaning.

Maeder \& Meynet
(2001) have recently reported a breakthrough on the theroetical front,
in terms of understanding the effect of rotation on the evolution of low
metallicity stars. It is hoped that improvements in the observational
arena will help best utilize such progress. 

\subsection{The Relative Number of RSGs and Wolf-Rayet Stars}

Maeder, Lequeux, \& Azzoapardi (1980) suggested that the relative
number of red supergiants and Wolf-Rayet stars should be a very sensitive
function of metallicity.  For some mass ranges, we expect that a massive
star evolves first to the RSG phase, and then to a W-R phase.  How quickly
a star becomes a W-R will depend upon the amount of mass-loss, and mass-loss
rates are lower (for a star of a given luminosity) at lower metallicities.
Thus the relative proportion of RSGs to W-Rs should decrease with increasing
metallicity.   Massey \& Johnson (1998) confirmed this trend among Local
Group galaxies, although the gradient with metallicity is roughly a factor of
20 less than that predicted by Maeder et al.\ (1980) on the basis of those
models. 

Massey \& Duffy (2001) have recently discovered 2 additional W-R stars in the SMC after an intense survey, establishing that there are not a significant
number of W-Rs still to be found in the SMC.  The population of W-R stars
in the LMC is also considered to be likely complete (Breysacher, Azzopardi, \& Testor 1999).
In Table~12 we compare the relative number of RSGs and W-Rs to those found in
other Local Group galaxies, where we now impose the same $M_{\rm bol}<-7.5$
restriction for the RSGs. In counting W-Rs in the LMC and SMC we chose to use 
only those detected by our survey, as the $B$ magnitudes of RSGs and W-Rs
are roughly compable (compare Tables~4 and 6 with Table~10), and thus would
have the similar selection problems.  Using all of the W-Rs instead would
lower the number ratio in the SMC to 6.9.  We also did not count any O-type
stars which appear in the LMC W-R catalog as W-Rs, e.g., any O3If*/WN6-A 
objects or the 30 Doradus ``super Of" stars (Massey \& Hunter 1998).

We illustrate the dramatic trend with metallicity in Fig.~\ref{fig:rsgswrs},
where we have used a logarithmic scaling to better show the tight
correlation. The relative number of RSGs and W-R star changes by
a factor of 160 from the SMC to M31.

\subsection{The Field Initial Mass Function, Revisited}

Massey et al.\ (1995)  discussed the stellar initial mass function
of the massive stars in the Clouds, and concluded that,  although the
OB associations all have an IMF slope which is indistinguishable from
Salpeter ($\Gamma=-1.35$), there is a substantial {\it field}
population of massive stars which has a considerably steeper slope
$\Gamma\sim-4$.  Many questions remain about the massive stars found
far from any OB association.  Were they born there?  Were they ejected
from neighboring associations?  In a number of cases the very early spectral
types (O3-O4) would seem to preclude these stars having immigrated, as the
short lifetimes of the O3 stage (1~Myr) places severe restrictions on how
far such a star could have traveled. (See additional discussion in Massey 1998a, 1999).

However, the Massey et al.\ (1995) study relied upon correcting the
number of stars in the H-R diagram for incompleteness; these corrections
became increasingly significant at lower masses, with only 25\% completeness
for the 25-40$\cal M_\odot$.  Although the vast majority of the stars in
the current sample lack spectroscopy, we are now photometrically complete
to 25$\cal M_\odot$.  We have used these data to recompute the IMF of the
{\it field} population of both the Large and Small Clouds. As with our
previous study, we {\it exclude} stars 
in or near any of the cataloged
OB associations.\footnote{For the current study we use a distance
cut-off of 2~arcmin from the boundary of an OB association; this
corresponds to 30~pc and 35~pc in the LMC and SMC, respectively.}
We count the number of stars in each mass bin, and then
normalize this number by the logarithm size of the mass bin, divide by
the average hydrogen-burning lifetime stars in the bin, and by the area in
kpc$^2$ to obtain Scalo's (1986) $\xi$.   We list these quantities in
Table~13. The assumption that we are making here
is that star-formation is ``steady state", i.e., that the number of stars
born is equal to the number of stellar deaths.  This is simply equivalent to
saying that over the past 10~Myr there have not been significant changes
in the star-formation rate {\it averaged over the entire galaxy.}   We again find for both the
Small and Large Cloud that the slope of this {\it field} population of
massive stars is $\sim 4$, very differenet from what is found within the
OB associations.  Incompleteness corrections therefore cannot be the reason,
unlike the explanation offered by Parker et al.\ (2001).

\acknowledgments

Support for this project was partially provided by four sources.  The project
was
begun while the author was a staff member at the National Optical Astronomy
Observatories.  
In so far as these data were needed for the
successful completion of our modeling of massive, hot stars in the Clouds,
partial support was also 
provided by NASA through grant number
8633 from the Space Telescope Science Institute, which is operated by
AURA, Inc., under NASA contract NAS5-26555.  Additional support was 
obtained from the Friends of Lowell Observatory.
By the time this paper was being prepared
for publication, the work was supported under NSF Grant AST-0093060.
We are grateful to CTIO for
the two enjoyable observing runs during which these data were obtained.
We are also grateful 
to Deidre A. Hunter for a critical reading of this manuscript,
nd for useuful comments by Nolan Walborn, Joel Parker, and an anonymous
referee.

\clearpage


\begin{figure}
\caption{The outlines of the eleven $1.2^\circ \times 1.2^\circ$ Schmidt
fields are
shown against this image of the LMC.  The image itself was obtained with
the ``parking lot camera" by Greg Bothun, and kindly made available to us
for the purposes of illustration. North is to the top and east is to the left.\label{fig:LMC}}
\end{figure}

\begin{figure}
\caption{The outlines of the six $1.2^\circ \times 1.2^\circ$ Schmidt 
are shown against this image of the SMC.  Like that of Fig.~1, this image
was obtained by Greg Bothun. North is to the top, and east is to the left.}
\end{figure}

\begin{figure}
\caption{Histograms showing the number of stars as a function of magnitude
for the LMC data.\label{fig:LMChists}}
\end{figure}

\begin{figure}
\caption{Histograms showing the number of stars as a function of magnitude
for the SMC data.\label{fig:SMChists}}
\end{figure}

\begin{figure}
\caption{The difference between our CCD photometry and previously publihsed
photoelectric (PE) photometry is shown.  \label{fig:photdifs}}
\end{figure}

\begin{figure}
\caption{Color-magnitude diagram of the LMC.  
In the top panel all the stars in 
Table~3A are plotted; in the middle panel every 10th point is shown.
The (de)reddening vector is shown by an arrow.
In the bottom panel we show the percentage of stars expected to
be foreground.\label{fig:lmccmd}}
\end{figure}

\begin{figure}
\caption{Color-magnitude diagram of the SMC.  
In the top panel all the stars in 
Table~3A are plotted; in the middle panel every tenth point is shown.
The (de)reddening vector is shown by an arrow.
In the bottom panel we show the percentage of stars expected to
be foreground; compare to the Fig.~6. \label{fig:smccmd}}
\end{figure}

\begin{figure}
\caption{The intrinsic
relation for dwarfs (solid curve) and supergiants (dashed curve) is shown,
based upon Massey (1998c) and FitzGerald (1970).
The small trapazoid shows where we expect to find
F5-F9 supergiants.  The data for the LMC and SMC has been corrected for
$E(B-V)=0.13$ and 0.09, respectively.
{\label{fig:2color}}}
\end{figure}

\begin{figure}
\caption{The reddest stars in the LMC and SMC are shown in the two-color
diagram $B-V$ vs. $V-R$.  The solid line shows the expectated
dividing line between red supergiants and foreground stars (Massey 1998b).  
The
filled circles denote spectroscopically confirmed supergiants.  
\label{fig:rsgs}}
\end{figure}

\begin{figure}
\caption{The H-R diagram of the LMC is shown, based upon the data presented
in Tables~8A and 10A.  The stellar evolutionary tracks of
Schaerer et al.\ (1993) are shown with solid lines,
corresponding to $z=0.008$. The filled circles denote stars with
slit spectroscopy. An upper limit of
$\log T_{\rm eff} =4.66$ was imposed for the stars with only
photometry. \label{fig:lmchrd}}
\end{figure}

\begin{figure}
\caption{The H-R diagram of the SMC is shown, based upon the data presented
in Tables~8B and 10B.  The stellar evolutionary tracks of
Schaller et al.\ (1992) are shown with solid lines,
corresponding to $z=0.001$. The filled circles denote stars with
slit spectroscopy. An upper limit of
$\log T_{\rm eff} =4.66$ was imposed for the stars with only
photometry.\label{fig:smchrd}}
\end{figure}

\begin{figure}
\caption{The relative number of luminous RSGs ($M_{\rm bol}<-7.5$) and 
W-R stars is shown as a function of the oxygen abundance.
\label{fig:rsgswrs}}
\end{figure}

 \begin{figure}
\caption{The stellar initial mass functions are show for the field stars
of the LMC and SMC. \label{fig:imf}}
\end{figure}
 
\clearpage
\begin{deluxetable}{l c c c c } 
\tablewidth{0pc}
\tablenum{1}
\tablecolumns{5}
\tablecaption{Color Terms and Extinction Coefficients}
\tablehead{
\colhead{Filter}
&\multicolumn{2}{c}{Color Terms ($c_3$)}
&\multicolumn{2}{c}{Extinction Coefficients ($c_2$)} \\
\colhead{}
&\colhead{1999}
&\colhead{2001}
&\colhead{1999}
&\colhead{2001}
}
\startdata
$U$ & $-$0.12 & $-$0.09 ($U-B>0.0$) &0.51 &0.35-0.63  \\
      &        & $-$0.19 ($U-B<0.0$)  \\
$B$ &  +0.09 & +0.07  & 0.35 & 0.23-0.31 \\
$V$ &+0.02 & +0.02  & 0.24 & 0.10-0.18 \\
$R$ &$-$0.02& $-$0.02  & 0.19& 0.07-0.14 \\
\enddata
\end{deluxetable}

\begin{deluxetable}{l c c c c c c c c c} 
\tablewidth{0pc}
\tablenum{2}
\tablecolumns{10}
\tablecaption{Average Statistical Errors}
\tablehead{
\colhead{Mag.}
&\multicolumn{4}{c}{LMC}
&
&\multicolumn{4}{c}{SMC} \\ \cline{2-5} \cline{7-10}
\colhead{}
&\colhead{$U$}
&\colhead{$B$}
&\colhead{$V$}
&\colhead{$R$} 
&
&\colhead{$U$}
&\colhead{$B$}
&\colhead{$V$}
&\colhead{$R$}
}
\startdata
12.0 & 0.01 & 0.01 & 0.01 & 0.01 && 0.01 & 0.01 & 0.01 & 0.01 \\
12.5 & 0.01 & 0.01 & 0.01 & 0.02 && 0.01 & 0.01 & 0.01 & 0.02 \\
13.0 & 0.01 & 0.02 & 0.01 & 0.02 && 0.01 & 0.01 & 0.01 & 0.02 \\
13.5 & 0.02 & 0.02 & 0.01 & 0.03 && 0.01 & 0.02 & 0.01 & 0.02 \\
14.0 & 0.02 & 0.02 & 0.02 & 0.04 && 0.02 & 0.02 & 0.02 & 0.03 \\
14.5 & 0.03 & 0.02 & 0.02 & 0.06 && 0.03 & 0.02 & 0.02 & 0.05 \\
15.0 & 0.04 & 0.03 & 0.04 & 0.09 && 0.04 & 0.03 & 0.04 & 0.07 \\ 
15.5 & 0.06 & 0.04 & 0.07 & 0.14 && 0.06 & 0.03 & 0.06 & 0.11 \\
16.0 & 0.09 & 0.05 & 0.10 & 0.21 && 0.08 & 0.04 & 0.08 & 0.18 \\
16.5 & 0.13 & 0.07 & 0.14 & 0.34 && 0.12 & 0.05 & 0.12 & 0.27 \\
17.0 & 0.19 & 0.08 & 0.21 & 0.56 && 0.17 & 0.08 & 0.18 & 0.44 \\
17.5 & 0.25 & 0.11 & 0.35 & 0.92 && 0.23 & 0.08 & 0.29 & 0.75 \\
\enddata
\end{deluxetable}

\clearpage
\begin{deluxetable}{r r r r r r r r r r r r r r } 
\tabletypesize{\scriptsize}
\tablewidth{0pc}
\tablenum{3A}
\tablecolumns{14}
\tablecaption{LMC Catalog\tablenotemark{a}}
\tablehead{
\colhead{LMC ID}
&\colhead{$\alpha_{2000}$}
&\colhead{$\delta_{2000}$}
&\colhead{$V$}
&\colhead{$B-V$}
&\colhead{$U-B$}\tablenotemark{b}
&\colhead{$V-R$}\tablenotemark{b}
&\colhead{$\sigma_V$}
&\colhead{$\sigma_{B-V}$}
&\colhead{$\sigma_{U-B}$}
&\colhead{$\sigma_{V-R}$}
&\colhead{$N_{BV}$}
&\colhead{$N_{U}$}
&\colhead{$N_{R}$} \\
}
\startdata
000001 &04 43 54.19 &$-$70 57 31.6 &14.51 & 1.04  &0.67 & 0.56 & 0.02 & 0.03  &0.08 & 0.04 & 1&  1&  1\\
000002 &04 43 57.78 &$-$70 53 35.7 &16.64 & 1.08 &99.99 & 0.93 & 0.12 & 0.15 & 0.00  &0.18  &1  &0  &1\\
000003 & 04 43 59.92 &$-$70 59 47.0& 13.30& 0.73 & 0.19 & 0.39 & 0.01 & 0.01 & 0.02  &0.02  &1  &1  &1\\
000004 &04 44 01.95 &$-$70 53 36.5 &16.95 & 1.13 &99.99 & 0.88 & 0.17 & 0.21 & 0.00  &0.25  &1  &0  &1\\
000005 &04 44 02.97 &$-$70 48 11.4 &17.01 & 0.85 &99.99 & 0.90 & 0.17 & 0.19 & 0.00  &0.25  &1  &0  &1\\
000006 &04 44 03.35 &$-$70 51 23.0 &17.00 & 0.78 &99.99 & 0.04 & 0.14 & 0.17 & 0.00  &0.41  &1  &0  &1\\
000007 &04 44 03.35 &$-$70 49 55.0 &16.27 & 0.94 &99.99 & 0.51 & 0.09 & 0.10 & 0.00  &0.16  &1  &0  &1\\
000008 &04 44 03.36 &$-$70 52 53.1 &17.02 & 0.28 & 0.57 &99.99 & 0.16 & 0.16 & 0.34  &0.00  &1  &1  &0\\
000009 &04 44 03.53 &$-$70 39 40.0 &16.91 &$-$0.18& $-$0.72& 99.99 & 0.15 & 0.14 & 0.09  &0.00  &1  &1 & 0\\
000010 &04 44 03.98 &$-$70 38 52.9 &16.22 & 1.20& 99.99 & 0.72 & 0.09 & 0.12 & 0.00 & 0.13  &1  &0  &1\\
\enddata
\tablenotetext{a}{The complete version of this table will be published electronically.}
\tablenotetext{b}{An entry of 99.99 for $U-B$ or $V-R$ indicates no measurement.}
\end{deluxetable}

\begin{deluxetable}{r r r r r r r r r r r r r r } 
\tabletypesize{\scriptsize}
\tablewidth{0pc}
\tablenum{3B}
\tablecolumns{14}
\tablecaption{SMC Catalog\tablenotemark{a}}
\tablehead{
\colhead{SMC ID}
&\colhead{$\alpha_{2000}$}
&\colhead{$\delta_{2000}$}
&\colhead{$V$}
&\colhead{$B-V$}
&\colhead{$U-B$}\tablenotemark{b}
&\colhead{$V-R$}\tablenotemark{b}
&\colhead{$\sigma_V$}
&\colhead{$\sigma_{B-V}$}
&\colhead{$\sigma_{U-B}$}
&\colhead{$\sigma_{V-R}$}
&\colhead{$N_{BV}$}
&\colhead{$N_{U}$}
&\colhead{$N_{R}$} \\
}
\startdata
000001 &00 40 00.49 &$-$74 01 57.8 &16.82 & 0.99 & 99.99  &0.83 & 0.13 & 0.14&  0.00 & 0.19  &1  &0 & 1\\
000002 &00 40 03.08 &$-$73 56 19.4 &16.30 &$-$0.07 &$-$0.67 &$-$0.18 & 0.08 & 0.08 & 0.05 & 0.27 & 1 & 1 & 1 \\
000003 &00 40 04.19 &$-$73 56 06.3 &15.88  &0.89  &0.83 & 0.72 & 0.06 & 0.06 & 0.20  &0.09  &1  &1  &1 \\
000004 &00 40 07.38 &$-$74 00 56.0 &16.73  &0.36  &0.24  &0.48 & 0.12 & 0.12 & 0.18 & 0.22 & 1 & 1 & 1 \\
000005 &00 40 07.59 &$-$73 51 27.6 &14.49  &0.61  &0.30 & 0.40 & 0.02 & 0.02 & 0.04 & 0.04 & 1  &1 & 1 \\
000006 &00 40 09.02 &$-$73 48 58.4 &16.74  &0.99 &99.99 & 0.61  &0.11 & 0.12 & 0.00 & 0.18 &  1 & 0 &  1\\
000007 &00 40 09.33 &$-$73 53 11.3 &17.19  &0.26 &0.12 &99.99  &0.18 & 0.18  &0.23  &0.00 & 1  & 1  & 0 \\
000008 &00 40 09.59 &$-$73 46 06.8 &15.95  &1.03 &99.99 & 0.56 & 0.06 & 0.07 & 0.00 & 0.10 & 1  & 0  & 1\\
000009 &00 40 10.89 &$-$73 59 26.0 &16.33  &0.20 &99.99 & 0.36 & 0.08 & 0.08 & 0.00 & 0.15 & 1 &  0 &  1 \\
000010 &00 40 11.22 &$-$73 55 03.9 &14.58  &0.66 &99.99 & 0.45 & 0.02 & 0.02&  0.00 & 0.03 & 1 &  0 &  1\\
\enddata
\tablenotetext{a}{The complete version of this table will be published electronically.}
\tablenotetext{b}{An entry of 99.99 for $U-B$ or $V-R$ indicates no measurement.}
\end{deluxetable}

\begin{deluxetable}{l c c c r r r r l l l}
\tabletypesize{\tiny}
\tablewidth{0pc}
\tablenum{4}
\tablecolumns{11}
\tablecaption{LMC Catalog Stars with Spectral Information}
\tablehead{
\colhead{Name\tablenotemark{a}}
&\colhead{Catalog Number}
&\colhead{$\alpha_{2000}$}
&\colhead{$\delta_{2000}$}
&\colhead{$V$}
&\colhead{$B-V$}
&\colhead{$U-B$}
&\colhead{$V-R$}
&\multicolumn{2}{c}{Spectral Data}
&\colhead{Other ID} \\ \cline{9-10}
&&&&&&&&Type & Source\tablenotemark{b} \\
}                                               
\startdata
\cutinhead{Stars from Rousseau et al.\ (1978)\tablenotemark{c}} \\
FDG   7     &0000119&04 44 28.55&-69 40 34.2&13.55& 0.01&-0.30& 0.12&A0: I     &OBJ 1   &               \\ 
FDG   8     &0000546&04 45 00.47&-69 53 25.1&14.01& 0.16& 0.08& 0.09&A5        &OBJ 1   &               \\ 
Sk -70   1  &0001015&04 45 32.43&-70 15 11.2&15.91&-0.23&-0.71& 0.19&WN3b        &SLIT 11 &BAT99-1         \\
Sk -69   1  &0001020&04 45 32.82&-69 28 01.5&13.39&-0.08&-0.97& 0.07&B1 III    &SLIT 7  &               \\ 
Sk -69   2  &0002243&04 46 54.97&-69 29 48.8&13.17&-0.14&-0.97&-0.01&B2:       &OBJ 2   &               \\ 
Sk -70   1A &0002790&04 47 27.69&-70 33 45.9&13.66&-0.06&-0.91& 0.08&O9 II     &SLIT 7  &               \\ 
Sk -69   2A &0002885&04 47 33.25&-69 14 33.0&12.53& 0.18&-0.06& 0.20&A3 I      &SLIT 4  &               \\
\enddata
\tablenotetext{a}{
For designations, please see the referenced sources.}

\tablenotetext{b}{``OBJ" indicates objective prism type, ``SLIT" indicates slit spectral
type.  References: (1)---Sanduleak 1969a;
(2)---Rousseau et al.\ 1978; 
(3)---Feast et al.\ 1960;
(4)---Ardenberg et al.\ 1972 and/or Brunet et al.\ 1973;
(5)---Walborn 1977;
(6)---Crampton 1979;
(7)---Conti, Garmany, \& Massey 1986;
(8)---Fitzpatrick 1988;
(9)---Massey et al.\ 1995;
(10)---Massey et al.\ 2000b;
(11)---Breysacher, Azzopardi, \& Testor 1999}

\tablenotetext{c}{We have included a few stars from the Rousseau et al.\ 1998
catalog that lack spectral types.}

\end{deluxetable}

\begin{deluxetable}{l c c c c c l l }
\tabletypesize{\scriptsize}
\tablewidth{0pc}
\tablenum{5}
\tablecolumns{8}
\tablecaption{Stars in Two Additional LMC Fields with Spectral Types }
\tablehead{
\colhead{Name\tablenotemark{a}}
&\colhead{$\alpha_{2000}$}
&\colhead{$\delta_{2000}$}
&\colhead{$V$}
&\colhead{$B-V$}
&\colhead{$U-B$}
&\colhead{Spectral Type}
&\colhead{Other ID} \\ 
}
\startdata                                
BI185     &05 30 40.18&-71 17 13.7& 14.25&-0.14&-0.94&O9V       &lmc1-  23 \\
lmc1-  94 &05 32 00.28&-71 14 08.4& 13.67& 0.05&-0.69&B1III     &          \\
lmc1- 170 &05 31 06.97&-71 10 47.7& 14.14&-0.17&-1.01&O8.5V     &          \\
lmc1- 212 &05 31 59.44&-71 09 42.1& 14.73&-0.18&-0.87&B1III     &          \\
lmc1- 226 &05 31 11.46&-71 09 29.5& 15.34&-0.17&-0.87&B1V       &          \\
lmc1- 233 &05 29 46.22&-71 09 21.8& 13.77&-0.02&-0.84&B0V       &          \\
lmc1- 246 &05 29 47.30&-71 09 00.4& 15.04& 0.06&-0.71&B1:V      &          \\
lmc1- 249 &05 29 55.20&-71 08 58.5& 14.95&-0.10&-0.82&B1V       &          \\
\enddata
\tablenotetext{a}{Designations are from Massey et al.\ 1995.}
\end{deluxetable}

\begin{deluxetable}{l c c c r r r r l l l}
\tabletypesize{\tiny}
\tablewidth{0pc}
\tablenum{6}
\tablecolumns{11}
\tablecaption{SMC Catalog Stars with Spectral Information}
\tablehead{
\colhead{Name}
&\colhead{Catalog Number}
&\colhead{$\alpha_{2000}$}
&\colhead{$\delta_{2000}$}
&\colhead{$V$}
&\colhead{$B-V$}
&\colhead{$U-B$}
&\colhead{$V-R$}
&\multicolumn{2}{c}{Spectral Data}
&\colhead{Other ID} \\ \cline{9-10}
&&&&&&&&Type & Source\tablenotemark{a} \\
}                                           
\startdata                   
\cutinhead{Azzopardi \& Vigneau (1982) Catalog} \\
AV   1A     &0001830&00 42 24.08&-73 16 51.2&13.94&-0.10&-0.85& 0.09&  B3      &OBJ 1   &               \\ 
AV   1      &0002325&00 42 53.33&-73 10 25.9&12.62& 0.11&-0.12& 0.10&  A0      &OBJ 1   &               \\ 
AV   2B     &0003459&00 43 49.95&-73 09 02.4&13.46&-0.14&-0.99&-0.02&  B0 IWW  &SLIT 8  &               \\ 
AV   3      &0004297&00 44 28.22&-72 58 05.0&13.93&-0.01&-0.60&-0.10&  B0      &OBJ 1   &               \\ 
AV   5      &0005127&00 45 05.86&-73 23 11.6&13.79&-0.02&-0.90&-0.04&  B1 I    &SLIT 8  &               \\ 
AV   6      &0005391&00 45 18.22&-73 15 23.0&13.31& 0.05&-0.91& 0.17&  B  pec  &SLIT 8  &               \\ 
AV   8      &0005768&00 45 35.45&-72 59 21.3&13.50&-0.04&-0.66&-0.06&  B8      &OBJ 1   &               \\
\enddata
\tablenotetext{a}{``OBJ" indicates objective prism type, ``SLIT" indicates slit spectral
type.  References: (1)---Azzopardi \& Vigneau 1982 and references therein; 
(2)---Feast et al.\ 1960;
(3)---Ardeberg \& Maurice 1977;
(4)---Dubois, Jaschek, \& Jaschek 1977;
(5)---Crampton \& Greasley 1982;
(6)---Humphreys 1983;
(7)---Walborn 1983;
(8)---Garmany et al.\ 1987;
(9)---Massey et al.\ 1995;
(10)---Massey et al.\ 2000b;
(11)---Sanduleak 1969b;
(12)---Massey, previously unpublished;
(13)---Massey \& Duffy 2001;
(14)---Walborn et al. 2000;
(15)---Walborn et al. 1995.
}      
\end{deluxetable}

\begin{deluxetable}{l l l c c c c c c c c c c c c c c}
\tabletypesize{\tiny}
\rotate
\tablewidth{0pt}
\tablenum{7}
\tablecolumns{17}
\setlength{\tabcolsep}{0.05in}
\tablecaption{Comparison of the Intrinsic $(U-B)_o$ with that Derived from
CCD and Photoelectric Photometry
}
\tablehead{
\colhead{Star}
&\colhead{Catalog}
&\colhead{Spectral}
&\multicolumn{2}{c}{FitzGerald(1970)}
&\colhead{ }
&\multicolumn{5}{c}{CCD Photometry}
&\colhead{ }
&\multicolumn{5}{c}{Photoelectric Photometry} \\ 
\cline{4-5} \cline{7-11} \cline{13-17} \\
\colhead{}
&\colhead{Number}
&\colhead{Type}
&\colhead{$(B-V)_o$}
&\colhead{$(U-B)_o$}
&\colhead{}
&\colhead{$(B-V)$}
&\colhead{$(U-B)$}
&\colhead{$E(B-V)$}
&\colhead{$(U-B)_o$}
&\colhead{$\Delta(U-B)_o$}
&\colhead{}
&\colhead{$(B-V)$}
&\colhead{$(U-B)$}
&\colhead{$E(B-V)$}
&\colhead{$(U-B)_o$}
&\colhead{$\Delta(U-B)_o$}  \\
}
\startdata
\cutinhead{SMC} \\
AV  90     &0019066 &A2 II:&0.05&-0.21&&  0.08&-0.13 & 0.03 &-0.15 &0.06&&   0.08& 0.12&0.03&0.10&0.22\\
AV 174     &0037700 &A5 I  &0.10&-0.09&&  0.23& 0.00 & 0.13 &-0.09 &0.00&&   0.23& 0.27&0.13&0.18&0.27\\
AV 286     &0052465 &A3 I  &0.06&-0.12&&  0.15&-0.02 & 0.09 &-0.08 &0.04&&   0.15& 0.32&0.09&0.26&0.38\\
AV 331     &0056294 &A2 I  &0.05&-0.26&&  0.12&-0.11 & 0.07 &-0.16 &0.10&&   0.10& 0.12&0.05&0.08&0.34\\
AV 463     &0072530 &A2 I  &0.03&-0.26&&  0.09&-0.17 & 0.06 &-0.21 &0.05&&   0.08& 0.11&0.05&0.07&0.33\\
\cutinhead{LMC} \\
Sk -69   2A&0002885 &A3 I  &0.06&-0.12&&  0.18&-0.06 & 0.09 &-0.12 &0.00&&   0.23& 0.20&0.17&0.08&0.20\\
\enddata
\end{deluxetable}

\begin{deluxetable}{l c c c c c c c c l}
\tabletypesize{\footnotesize}
\tablewidth{0pt}
\tablenum{8A}
\tablecolumns{10}
\tablecaption{LMC Blue Stars with $(B-V)_o<0.14$ and $M_{bol}<-7.0$\tablenotemark{a}}
\tablehead{
\colhead{Star}
&\colhead{$\alpha_{2000}$}
&\colhead{$\delta_{2000}$}
&\colhead{$V$}
&\colhead{$B-V$}
&\colhead{$U-B$}
&\colhead{$Q$}
&\colhead{$T_{\rm eff}$}
&\colhead{$M_{\rm bol}$}
&\colhead{Spectral Type}
}
\startdata       
000204&04 44 36.23&-69 28 22.2& 14.85&-0.02&-0.85&-0.84&4.485& -7.03&         \\
000277&04 44 42.10&-69 02 52.2& 15.85&-0.02&-0.96&-0.95&4.628& -7.01&         \\
000322&04 44 45.45&-69 01 23.8& 14.59&-0.01&-1.02&-1.01&4.660& -8.49&         \\
000695&04 45 11.81&-69 22 34.1& 15.78&-0.42&99.99&99.99&4.660& -7.30&         \\
000907&04 45 25.33&-69 35 14.4& 15.75&-0.19&-0.70&-0.56&4.639& -7.19&         \\
000936&04 45 27.41&-69 00 33.6& 14.24&-0.12&-0.93&-0.84&4.495& -7.71&         \\
001015&04 45 32.43&-70 15 11.2& 15.91&-0.23&-0.71&-0.54&4.660& -7.17&         \\
001020&04 45 32.82&-69 28 01.5& 13.39&-0.08&-0.97&-0.91&4.372& -7.72&B1 III   \\
001485&04 46 05.28&-68 46 50.7& 13.83&-0.12&-0.88&-0.79&4.396& -7.44&         \\
001875&04 46 30.57&-69 47 01.3& 15.85&-0.24&-0.32&-0.15&4.660& -7.23&         \\
\enddata
\tablenotetext{a}{The complete version of this table will be published electronically.}
\end{deluxetable}        

\begin{deluxetable}{l c c c c c c c c l}
\tabletypesize{\footnotesize}
\tablewidth{0pt}
\tablenum{8B}
\tablecolumns{10}
\tablecaption{SMC Blue Stars with $(B-V)_o<0.14$ and $M_{bol}<-7.0$\tablenotemark{a}}
\tablehead{
\colhead{Star}
&\colhead{$\alpha_{2000}$}
&\colhead{$\delta_{2000}$}
&\colhead{$V$}
&\colhead{$B-V$}
&\colhead{$U-B$}
&\colhead{$Q$}
&\colhead{$T_{\rm eff}$}
&\colhead{$M_{\rm bol}$}
&\colhead{Spectral Type}
}
\startdata     
000107&00 40 28.11&-73 32 32.9& 15.00&-0.04&-0.95&-0.92&4.594& -8.00&         \\
000259&00 40 41.66&-73 22 29.8& 16.44&-0.31&-0.13& 0.09&4.660& -7.01&         \\
000298&00 40 43.96&-73 24 22.8& 15.12& 0.06&-0.78&-0.82&4.471& -7.04&         \\
000314&00 40 45.24&-73 22 56.6& 14.39&-0.15&-0.46&-0.35&4.377& -7.13&         \\
000515&00 40 59.94&-73 20 35.9& 15.85&-0.38&-0.51&-0.24&4.660& -7.60&         \\
000665&00 41 11.26&-73 21 39.1& 16.30&-0.26&-0.72&-0.53&4.660& -7.15&         \\
000930&00 41 26.92&-73 26 30.9& 15.19& 0.09&-0.86&-0.92&4.599& -7.85&         \\
001076&00 41 35.75&-73 29 17.5& 15.06&-0.14&-0.92&-0.82&4.466& -7.07&         \\
001148&00 41 41.25&-73 32 45.8& 15.92&-0.23&-0.24&-0.07&4.639& -7.39&         \\
003459&00 43 49.95&-73 09 02.4& 13.46&-0.14&-0.99&-0.89&4.460& -8.63&  B0 IWW \\
\enddata
\tablenotetext{a}{The complete version of this table will be published electronically.}
\end{deluxetable}

\begin{deluxetable}{l c c c c c c l } 
\tabletypesize{\footnotesize}
\tablewidth{0pt}
\tablenum{9A}
\tablecolumns{8}
\tablecaption{Potential F Supergiants in the LMC}
\tablehead{
\colhead{Star}
&\colhead{$\alpha_{2000}$}
&\colhead{$\delta_{2000}$}
&\colhead{$V$}
&\colhead{$B-V$}
&\colhead{$U-B$}
&\colhead{$V-R$}
}
\startdata
000098&04 44 26.30&-69 45 12.2&14.05& 0.81& 0.46& 0.53&     \\        
001127&04 45 41.27&-69 49 08.6&13.37& 0.80& 0.49& 0.47&     \\        
001634&04 46 14.82&-70 27 06.2&13.96& 0.80& 0.44& 0.49&     \\        
002297&04 46 57.91&-69 30 51.0&14.10& 0.75& 0.49& 0.46&     \\        
002991&04 47 39.52&-68 45 34.1&13.42& 0.79& 0.51& 0.52&     \\        
003155&04 47 49.71&-69 45 08.6&13.55& 0.81& 0.45& 0.48&     \\        
004653&04 49 01.35&-68 58 36.9&13.57& 0.68& 0.40& 0.48&     \\        
004970&04 49 11.90&-69 36 38.7&11.80& 0.76& 0.44& 0.48&     \\
\enddata
\end{deluxetable}

\begin{deluxetable}{l c c c c c c l } 
\tabletypesize{\normalsize}
\tablewidth{0pt}
\tablenum{9B}
\tablecolumns{8}
\tablecaption{Potential F Supergiants in the SMC}
\tablehead{
\colhead{Star}
&\colhead{$\alpha_{2000}$}
&\colhead{$\delta_{2000}$}
&\colhead{$V$}
&\colhead{$B-V$}
&\colhead{$U-B$}
&\colhead{$V-R$}
}
\startdata
006211&00 45 54.08&-72 52 22.7&14.32& 0.76& 0.47& 0.49&                       \\
009160&00 47 44.60&-74 02 33.5&14.30& 0.76& 0.44& 0.51&                       \\
010516&00 48 19.20&-73 55 17.2&13.55& 0.73& 0.42& 0.40&                       \\
011348&00 48 37.94&-73 41 24.2&13.98& 0.75& 0.41& 0.46&                       \\
012116&00 48 55.90&-73 43 44.5&14.20& 0.70& 0.39& 0.42&                       \\
013014&00 49 14.85&-73 40 00.7&13.58& 0.77& 0.49& 0.50&                       \\
014271&00 49 41.60&-73 26 40.7&12.98& 0.67& 0.41& 0.48&                       \\
015240&00 50 01.20&-72 45 16.0&14.33& 0.73& 0.59&99.99&                       \\
\enddata
\end{deluxetable}

\begin{deluxetable}{l c c c c c c c c l l}
\tabletypesize{\footnotesize}
\tablewidth{0pt}
\tablenum{10A}
\tablecolumns{11}
\tablecaption{LMC Red Stars with $(V-R)_o>0.6$ and $M_{bol}<-7.0$}
\tablehead{
\colhead{Star}
&\colhead{$\alpha_{2000}$}
&\colhead{$\delta_{2000}$}
&\colhead{$V$}
&\colhead{$B-V$}
&\colhead{$V-R$}
&\colhead{$\Delta_{B-V}$\tablenotemark{a}}
&\colhead{$T_{\rm eff}$}
&\colhead{$M_{\rm bol}$}
&\colhead{Spectral Type\tablenotemark{b}}
&\colhead{Id\tablenotemark{c}}
}
\startdata                             
000202&04 44 36.17&-70 43 02.7& 13.45& 1.64& 1.52& 0.02&3.490& -3.09&         &         \\
000298&04 44 43.66&-70 16 05.2& 11.63& 1.22& 0.67& 0.18&3.653& -4.28&         &         \\
000311&04 44 44.82&-69 05 47.4& 12.26& 1.57& 0.94& 0.09&3.580& -4.18&         &         \\
000415&04 44 51.49&-70 26 41.0& 12.78& 1.50& 0.94& 0.02&3.580& -2.58&         &         \\
000570&04 45 01.92&-70 27 55.6& 12.93& 1.46& 0.99&-0.07&3.569& -3.73&         &         \\
002634&04 47 18.55&-69 18 29.3& 11.55& 1.22& 0.69& 0.14&3.647& -3.57&         &         \\
003533&04 48 07.44&-69 22 32.2& 12.30& 1.59& 0.92& 0.14&3.585& -4.29&         &         \\
003677&04 48 16.66&-69 57 12.5& 11.14& 1.34& 0.72& 0.20&3.638& -2.43&         &         \\
\enddata
\tablenotetext{a}{$\Delta_{B-V}=(B-V)-(B-V)_{\rm cut}$, where
$(B-V)_{\rm cut}$ is the expected dividing line between RSGs and 
foreground stars.  Stars with positive values
are expected to be red supergiants; highly negative values suggest foreground
stars.}       
\tablenotetext{b}{Spectral types from Humphreys (1979).}
\tablenotetext{c}{Identification for stars with spectral types from
Sanduleak \& Philip (1977).}
\end{deluxetable}

\begin{deluxetable}{l c c c c c c c c l l }
\tabletypesize{\footnotesize}
\tablewidth{0pt}
\tablenum{10B}
\tablecolumns{11}
\tablecaption{SMC Red Stars with $(V-R)_o>0.6$ and $M_{bol}<-7.0$}
\tablehead{
\colhead{Star}
&\colhead{$\alpha_{2000}$}
&\colhead{$\delta_{2000}$}
&\colhead{$V$}
&\colhead{$B-V$}
&\colhead{$V-R$}
&\colhead{$\Delta_{B-V}$\tablenotemark{a}}
&\colhead{$T_{\rm eff}$}
&\colhead{$M_{\rm bol}$}
&\colhead{Spectral Type\tablenotemark{b}}
&\colhead{Id\tablenotemark{c}}
}
\startdata                       
001152&00 41 41.52&-73 34 31.2& 11.17& 1.32& 0.69& 0.24&3.641& -3.19&         &         \\
003196&00 43 37.94&-73 20 12.6& 13.11& 1.67& 0.87& 0.28&3.592& -4.96&         &         \\
005092&00 45 04.56&-73 05 27.4& 12.90& 2.03& 1.18& 0.36&3.530& -2.93&M1 I     &SkKM13   \\
005315&00 45 14.59&-73 16 17.6& 15.72& 0.85& 2.57& 1.71&3.442& -7.54&         &         \\
008324&00 47 16.84&-73 08 08.4& 13.08& 1.64& 0.85& 0.28&3.597& -3.69&         &         \\
008367&00 47 18.11&-73 10 39.3& 12.46& 1.40& 0.93&-0.06&3.578& -6.55&         &         \\
008416&00 47 19.80&-73 10 40.5& 12.51& 1.51& 0.91& 0.07&3.582& -4.12&         &         \\
008930&00 47 36.94&-73 04 44.3& 12.68& 2.00& 1.06& 0.41&3.551& -4.28&M1 Ia    &SkKM22   \\
\enddata
\tablenotetext{a}{$\Delta_{B-V}=(B-V)-(B-V)_{\rm cut}$, where $B-V)_{\rm cut}$
is the expected dividing line between RSGs and foreground stars. 
Stars with positive values
are expected to be red supergiants; highly negative values suggest foreground
stars.}
\tablenotetext{b}{Spectral types from Elias, Frogel, \& Humphreys (1985).}
\tablenotetext{c}{Identification for stars with spectral types from
Sanduleak (1989).}
\tablenotetext{d}{R20 was classified K5 Ia by Feast et al.\ (1960).}
\end{deluxetable}

\begin{deluxetable}{l r r r r r }
\tablewidth{0pc}
\tablenum{11}
\tablecolumns{6}
\tablecaption{B/R Ratios in the LMC and SMC\tablenotemark{a}}
\tablehead{
\colhead{}
&\multicolumn{3}{c}{Number of Stars} \\ \cline{2-4}
\colhead{Sample}
&\colhead{Blue}
&\colhead{Red}
&\colhead{RSGs\tablenotemark{b}}
&\colhead{$B/R$}
&\colhead{$B/R_{\rm corr}$\tablenotemark{c}} \\
}
\startdata
LMC  (M) & 3134  &  69 &  67  &     45 &   47 \\
LMC (K+M)& 3134  & 134 & 127  &     23 &   25 \\
SMC  (M) & 1489  &  12 &   9  &    125 &   165 \\
SMC (K+M)& 1489  &  81 &  76  &     18 &   20 \\
\enddata
\tablenotetext{a}
{Using $M_{\rm bol} < -7.5$.  ``Blue stars" 
include all stars with $(B-V)_o<0.14$,
roughly corresdponding to A9~I.  The ``Red stars" include all stars
with $(V-R)_o>0.60$ for (K+M), or $(V-R)_o>1.0$ for M.} 
\tablenotetext{b}{The ``RSG" column is the number of red stars corrected
by eliminating suspected foreground stars based upon a value of 
$\Delta_{B-V}<-0.10$
in Table~10.}
\tablenotetext{c}{The corrected blue-to-red ratio  $B/R_{\rm corr}$ is
computed using the number of blue stars and the entry under ``RSG".}
\end{deluxetable}

\begin{deluxetable}{l c c r r c }
\tablewidth{0pc}
\tablenum{12}
\tablecolumns{6}
\tablecaption{The Relative Number of Red Supergiants and Wolf-Rayet Stars}
\tablehead{
\colhead{}
&\multicolumn{2}{c}{Abundance} \\ \cline{2-3}
\colhead{Galaxy}
&\colhead{12+log(O/H)}
&\colhead{Reference\tablenotemark{a}}
&\colhead{RSGs\tablenotemark{b}}
&\colhead{W-Rs\tablenotemark{c}}
&\colhead{RSG/W-R\tablenotemark{d}} \\
}
\startdata
SMC                            & 8.13 & 1 & 76  & 9   & $8.4\pm3$    \\
LMC                            & 8.37 & 1 & 127 & 62  & $2.0\pm0.3$  \\
NGC6822                        & 8.25 & 2 &  8  &  4   & $2.0\pm 1.2$ \\
M33 inner ($\rho<0.25$)        & 8.70 & 3 &  3 &  17   & $0.18\pm0.11$ \\
M33 middle ($0.25<\rho\ge0.5$) & 8.60 & 3 &  7 &  22   & $0.32\pm0.14$ \\
M33 outer ($\rho>0.5$)         & 8.35 & 3 & 26 & 18   & $1.4\pm0.4$   \\
M31                            & 9.00 & 4 & 1 &  19   & $0.053\pm0.053$ \\
\enddata
\tablenotetext{a}{References--(1) Russell \& Dopita 1990; (2) Pagel, Edmunds, \& Smith 1980; (3) Garnett et al.\ 1997; (4) Zaritzky et al.\ 1994.}
\tablenotetext{b}{$M_{\rm bol}<-7.5$.}
\tablenotetext{c}{From Massey \& Johnson 1998, Table 7.}
\tablenotetext{d}{The errors quoted are based upon $\sqrt{N}$ stastistics.}
\end{deluxetable}

\begin{deluxetable}{l r r r c }
\tablewidth{0pt}
\tablenum{13}
\tablecolumns{5}
\tablecaption{Initial Mass Function Outside OB Associations}
\tablehead{
\colhead{}
&\multicolumn{4}{c}{Mass Bin ($\cal M_\odot$)} \\ \cline{2-5}
\colhead{Parameter}
&\colhead{120-85}
&\colhead{85-60}
&\colhead{60-40}
&\colhead{40-25}
}
\startdata
\cutinhead{LMC}
Number of stars           &   1    &  7    &   69  & 396  \\
$\log \xi$ [Myr$^{-1}$ kpc$^{2}$] & -0.65 & 0.19 & 0.99 & 1.44 \\
$\Gamma$ &\multicolumn{4}{c}{$-4.2\pm0.5$} \\
\cutinhead{SMC}
Number of stars           &   1    &  4    &   46  & 238  \\
$\log \xi$ [Myr$^{-1}$ kpc$^{2}$] & -0.60 & -0.08 & 0.86 & 1.29 \\
$\Gamma$ & \multicolumn{4}{c}{$-4.0\pm0.5$} \\
\enddata
\end{deluxetable}


\begin{references}

\reference {} Ardeberg, A. Brunet, J. P., Maurice, E., \& Prevot, L. 1972, A\&AS, 6, 249

\reference {} Ardeberg, A., \& Maurice, E. 1977, A\&AS, 30, 261

\reference {} Azzopardi, M., \& Vigneau, J. 1975, A\&AS, 19, 271

\reference {} Azzopardi, M., \& Vigneau, J. 1979, A\&A, 35, 353

\reference {} Azzopardi, M., \& Vigneau, J. 1982, A\&AS, 50, 291

\reference {} Bahcall, J. N., \& Soneira, R. M. 1980, ApJS, 44, 73

\reference {} Berkhuijsen, E. M., Humphreys, R. M., Ghigo, F. D., \& Zumach, W.
1988, A\&AS, 76, 65

\reference {} Breysacher, J., Azzopardi, M., \& Testor, G. 1999, A\&AS, 137, 117

\reference {} Brunet, J. P., Imbert, M., Martin, N., Mianes, P., Prevot,
L., Rebeirot, E., \& Rousseau, J. 1975, A\&AS, 21, 109

\reference {} Brunet, J. P., Maurice, E., Muratorio, G., \& Prevot, L. 1973, A\&AS, 9, 447

\reference {} Brunish, W. M., Gallagher, J. S., Truran, J. W. 1986, AJ, 91, 598

\reference {} Cioni, M.-R. et al.\ 2000, A\&AS, 144, 235

\reference {} Conti, P. S., Garmany, C. D., \& Massey, P.  1986, AJ, 92, 48

\reference {} Crampton, D. C. 1979, ApJ, 230, 717

\reference {} Crampton, D. C., \& Greasley, J. 1982, PASP, 94, 31

\reference {} Dubois, P., Jascheck, M., \& Jaschek, C. 1977, A\&A, 60, 205

\reference {} Elias, J. H., Frogel, J. A., \& Humphreys, R. M. 1985, ApJS, 57, 
91

\reference {} Feast, M. W., Thackeray, A. D., \& Wesselink, A. J. 1960, MNRAS, 121, 337

\reference {} FitzGerald, M. P. 1970, A\&A, 4, 234

\reference {} Fitzpatrick, E. L. 1988, ApJ, 335, 703

\reference {} Fitzpatrick, E. L., \& Garmany, C. D. 1990, ApJ, 363, 119

\reference {} Garmany, C. D., Conti, P. S., \& Massey, P. 1987, AJ,
93, 1070 

\reference {} Garnett, D. R., Shields, G. A., Skillman, E. D., Sagan, S. P.,
\& Dufour, R. J. 1997, ApJ, 489, 63

\reference {} Humphreys, R. M. 1979, ApJS, 39, 389

\reference {} Humphreys, R. M. 1983, ApJ, 265, 176

\reference {} Humphreys, R. M, \& McElroy, D. B. 1984, ApJ, 284, 565

\reference {} Humphreys, R. M, \& Sandage, A. R. 1980, ApJS, 44, 319

\reference {} Isserstedt, J. 1975, A\&AS, 19, 259

\reference {} Isserstedt, J. 1978, A\&AS, 33, 193

\reference {} Isserstedt, J. 1979, A\&AS, 38, 239

\reference {} Isserstedt, J. 1982, A\&AS, 50, 7

\reference {} Ivanov, G. R., Freedman, W. L., \& Madore, B. F. 1993, ApJS, 89, 85

\reference {} Kurucz, R. 1992, in The Stellar Populations of Galaxies, ed.
B. Barbuy \& A. Renzini (Dodrecht: Kluwer), 225

\reference {} Landolt, A. U. 1992, AJ, 104, 340

\reference {} Langer, N., \& Maeder, A. 1995, A\&A, 295, 685

\reference {} Maeder, A., Lequeux, J., \& Azzopadi, M. 1980, A\&A, 189, 34

\reference {} Maeder, A., \& Meynet, G. 2001, A\&A, 373, 555

\reference {} Massey, P. 1998a, in The Stellar Initial Mass Function, 38th Herstmonceux Conference, ed.\ G. Gilmore \& D. Howell (San Francisco, ASP), 17

\reference {} Massey, P. 1998b, ApJ, 501, 153

\reference {} Massey, P. 1998c, in Stellar Astrophysics for the Local Group,
ed.\ A. Aparicio, A. Herrero, \& F. Sanchez 
(Cambridge, Cambridge Univ.\ Press), 95

\reference {} Massey, P. 1999, in New Views of the Magellanic Clouds,
ed. Y-H Chu, N. B. Suntzeff, J. E. Hesser, \& D. A. Bohlender
(Provo, ASP), 173

\reference {} Massey, P., Armandroff, T., DeVeny, J., Claver, C.,
Harmer, C., Jacoby, G., Schoening, B., \& Silva, D., 2000a, Direct Imaging
Manual for Kitt Peak (Tucson, NOAO)

\reference {} Massey, P., \& Duffy, A. 2001, ApJ, 550, 713

\reference {} Massey, P., \& Hunter, D. A. 1998, ApJ, 493, 180

\reference {} Massey, P., \& Johnson, O. 1998, ApJ, 505, 793

\reference {} Massey, P., Lang, C. C., DeGioia-Eastwood, K., \& Garmany, C. D. 
1995, ApJ, 438, 188

\reference {} Massey, P., Waterhouse, E., \& DeGioia-Eastwood 2000b, AJ,
119, 2214

\reference {} Oestreicher, M. O. \& Schmidt-Kaler, Th. 1999, 
Astron. Nachr. 320, 6, 385

\reference {} Pagel, B. E. J., Edmunds, M. G., \& Smith, G. 1980, MNRAS, 193, 219

\reference {} Parker, J. W., Hill, J. K., Cornet, R. H., Hollis, J.,
Zamkoff, E., Bohlin, R. C., O'Connell, R. W., Neff, S. G., Roberts,
M. S., Smith, A. M., \& Stecher, T. P. 1998, AJ, 116, 180

\reference {} Parker, J. W., Zaritsky, D., Stecher, T. P., Harris, J., \& Massey, P. 2001, AJ, 121, 891

\reference {} Prevot, L., Martin, N., Maurice, E., Rebeirot, E., \& Rousseau, J.1983, A\&AS, 53, 255

\reference {} Rousseau, J., Martin, N., Prevot, L., Rebeirot, E., Robin, A., \& Brunet, J. P. 1978, A\&AS, 31, 243

\reference {} Russell, S. C. \& Dopita, M. A. 1990, ApJS, 74, 93

\reference {} Sanduleak, N. 1968, AJ, 73, 246

\reference {} Sanduleak, N. 1969a, Contr.\ Cerro Tololo Inter-American Obs., 89

\reference {} Sanduleak, N. 1969b, AJ, 74, 877

\reference {} Sanduleak, N. 1975, A\&A, 39, 461

\reference {} Sanduleak, N. 1989, AJ, 98, 825

\reference {} Sanduleak, N., \& Philip, A. G. 1977, Publ.\ Warner and Swasey
Obs., 2, 105

\reference {} Scalo, J. M. 1986, Fund. Cosmic Phys., 11, 1

\reference {} Schaerer, D., Meynet, G., Maeder, A., \& Schaller, G. 1993,
A\&AS, 98, 523

\reference {} Schaller, G., Schaerer, D., Meynet, G., \& Maeder, A. 1992,
A\&AS, 96, 269

\reference {} Smith, R. C. \& the MCELS team 1999, 
in New Views of the Magellanic
Clouds, ed. Y.-H. Chu, N. B. Suntzeff, J. E. Hesser, \& D. A Bohlender
(Provo, ASP), 28

\reference {} Thackeray, A. D. 1962, Obs, 82, 207

\reference {} Vacca, W. D., Garmany, C. D., \& Shull, J. M. 1996, ApJ, 460, 914


\reference {} van den Bergh, S. 1973, ApJ, 183, L123

\reference {} van den Bergh, S. 2000, The Galaxies of the Local Group (Cambridge: Cambridge Univ. Press)

\reference {} Van Dyk, S. D., \& Cutri, R. 1999, in New Views of the Magellanic
Clouds, ed. Y.-H. Chu, N. B. Suntzeff, J. E. Hesser, \& D. A Bohlender
(Provo, ASP), 363


\reference {} Walborn, N. R. 1977, ApJ,  215, 53

\reference {} Walborn, N. R. 1983, ApJ, 265, 716

\reference {} Walborn, N. R., Lennon, D. J., Haser, S. M., Kudritzki, R. P.,
\& Voels, S. A. 1995, PASP, 107, 104

\reference {} Walborn, N. R., Lennon, D. J., Heap, S. R., Linder, D. J.,
Smith, L. J., Evans, C. J., \& Parker, J. W. 2000, PASP, 112, 1243

\reference {} Westerlund, B. E. 1997, The Magellanic Clouds (Cambridge: Cambridge Univ. Press)

\reference {} Zaritsky, D., Harris, J., \& Thompson, I. 1997, AJ, 114, 1002

\reference {} Zaritsky, D., Grebel, E. K., Harris, J., \& Thompson, I. 1999, 
in New Views of the Magellanic
Clouds, ed. Y.-H. Chu, N. B. Suntzeff, J. E. Hesser, \& D. A Bohlender
(Provo, ASP), 320

\reference {} Zaritsky, D., Kennicutt, R. C., \& Huchra, J. P. 1994, ApJ, 
420, 87

\end{references}
\end{document}